\documentclass[aps,twocolumn,superscriptaddress,showpacs]{revtex4-1}
\usepackage[dvips]{graphicx}
\usepackage{latexsym}
\usepackage{amsmath}
\usepackage{amsfonts}
\usepackage{amssymb}
\usepackage{bm}
\usepackage{color}
\begin{document}
\newcommand{\fig}[2]{\includegraphics[width=#1]{#2}}
\newcommand{{\vhf}}{\chi^\text{v}_f}
\newcommand{{\vhd}}{\chi^\text{v}_d}
\newcommand{{\vpd}}{\Delta^\text{v}_d}
\newcommand{{\ved}}{\epsilon^\text{v}_d}
\newcommand{{\vved}}{\varepsilon^\text{v}_d}
\newcommand{\la}{\langle}
\newcommand{\ra}{\rangle}
\newcommand{\dg}{\dagger}
\newcommand{\upa}{\uparrow}
\newcommand{\dna}{\downarrow}
\newcommand{\as}{{\alpha\sigma}}
\newcommand{\hH}{{\hat{\mathcal{H}}}}
\newcommand{\hn}{{\hat{n}}}
\newcommand{\hP}{{\hat{P}}}
\newcommand{{\bk}}{{\bf k}}
\newcommand{{\bq}}{{\bf q}}
\newcommand{{\tr}}{{\rm tr}}
\newcommand{\pprl}{Phys. Rev. Lett. \ }
\newcommand{\pprb}{Phys. Rev. {B}}
\newcommand{\dx}{$d_{x^2}$\ }
\newcommand{\dz}{$d_{z^2}$\ }

\title{Nodeless high-T$_c$ superconductivity in highly-overdoped monolayer CuO$_2$}

\author{Kun Jiang}
\affiliation{Department of Physics, Boston College, Chestnut Hill, MA 02467, USA}
\affiliation{Beijing National Laboratory for Condensed Matter Physics and Institute of Physics,
	Chinese Academy of Sciences, Beijing 100190, China}
\author{Xianxin Wu}
\affiliation{Beijing National Laboratory for Condensed Matter Physics and Institute of Physics,
	Chinese Academy of Sciences, Beijing 100190, China}
\affiliation{Institut f\"ur Theoretische Physik und Astrophysik,
	Julius-Maximilians-Universit\"at W\"urzburg, 97074 W\"urzburg, Germany}
\author{Jiangping Hu}
\affiliation{Beijing National Laboratory for Condensed Matter Physics and Institute of Physics,
	Chinese Academy of Sciences, Beijing 100190, China}
\affiliation{Collaborative Innovation Center of Quantum Matter,
Beijing, China}
\affiliation{Kavli Institute of Theoretical Sciences, University of Chinese Academy of Sciences,
Beijing, 100190, China}
\author{Ziqiang Wang}
\affiliation{Department of Physics, Boston College, Chestnut Hill, MA 02467, USA}

\date{\today}

\begin{abstract}

We study the electronic structure and superconductivity in CuO$_2$ monolayer grown recently on $d$-wave cuprate superconductor Bi$_2$Sr$_2$CaCu$_2$O$_{8+\delta}$. Density functional theory calculations indicate significant charge transfer across the interface such that the CuO$_2$ monolayer is heavily overdoped into the hole-rich regime yet inaccessible in bulk cuprates. We show that both the Cu $d_{x^2-y^2}$ and $d_{3z^2-r^2}$ orbitals become important and the Fermi surface contains one electron and one hole pocket associated with the two orbitals respectively. Constructing a minimal correlated two-orbital model for the $e_g$ complex,
we show that the spin-orbital exchange interactions produce a nodeless superconductor with extended $s$-wave pairing symmetry and a pairing energy gap comparable to the bulk $d$-wave gap, in agreement with recent experiments. The findings point to a direction of realizing new high-$T_c$ superconductors
in ozone grown transition-metal-oxide monolayer heterostructures.

\typeout{polish abstract}
\end{abstract}


\maketitle

The commonly held belief of the high-$T_c$ cuprate superconducting (SC) state \cite{bednorz} is that the superconductivity originates from two-dimensional copper-oxide (CuO$_2$) planes with a nodal $d$-wave pairing symmetry \cite{tsuei_prl,tsuei_rmp,lee}. In a recent attempt to directly probe the SC state in the copper-oxide plane, monolayer
CuO$_2$ on Bi$_2$Sr$_2$CaCu$_2$O$_{8+\delta}$ (Bi2212) has been grown successfully by the state-of-the-art ozone molecular beam epitaxy (MBE) \cite{xue}.
In contrast to the widely observed V-shaped local density of states (LDOS) typical of a nodal $d$-wave pairing gap in bulk cuprates, scanning tunneling microscopy (STM) on the CuO$_2$ monolayer reveals a robust U-shaped LDOS characteristic of a nodeless SC gap, which is further shown to be insensitive to nonmagnetic impurities \cite{xue}. Several theoretical scenarios have been proposed for this remarkable observation, largely based on the SC proximity effect
but with the $d$-wave nodes avoided by different Fermi surfaces or coexisting magnetism in the monolayer \cite{zhang1,zhang2,chen,fawang}. Here, we propose a different scenario. We argue that the CuO$_2$ monolayer has a new electronic structure due to interface charge transfer
and exhibits an intrinsic nodeless, $s$-wave SC state. Thus while the monolayer may not be representative of the bulk CuO$_2$ layers, it has potentially realized the direction of finding new and novel form of high-T$_c$ superconductors in transition-metal-oxide heterostructures by interface charge transfer.

The main findings are summarized in the schematic phase diagram
shown in Fig.~1. The left side of Fig.~1 has been realized by hole-doping the antiferromagnetic (AF) parent state in bulk cuprates, where the 3$d^9$ Cu$^{2+}$ has three $e_g$ electrons ($n_e=3$) occupy the well-split $d_{x^2-y^2}$ ($d_{x^2}$) and $d_{3z^2-r^2}$ ($d_{z^2}$) orbitals due to
Jahn-Teller distortion.
The $d$-wave superconductor emerges under the SC dome with a maximum $T_c$ around an optimal doping concentration $x_h\sim0.16$. Experiments show that the metallic state in bulk cuprates has a single band of the \dx character \cite{zhang_rice} crossing the Fermi level. Note that heavy overdoping is difficult and the region with $x_h>0.3$ has not been accessible in bulk cuprates.
\begin{figure}
	\begin{center}
		\fig{3.4in}{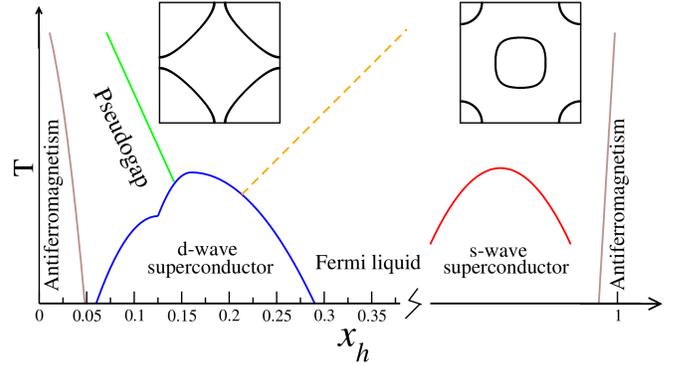}\caption{Schematic phase diagram as a function of hole doping $x_h$, contrasting the single-band $d$-wave SC phase realized in bulk cuprates (left side) with the two-orbital nodeless SC phase in the hole-rich CuO$_2$/Bi2212 monolayer (right side). The corresponding FS is shown in insets.}
	\end{center}
	\vskip-0.5cm
\end{figure}

The right side of Fig.~1 is conjectured for the monolayer CuO$_2$/Bi2212. Based on the experimental evidence suggesting that the monolayer crystalizes into CuO$_2$ \cite{xue}, significant charge transfer must occur between the CuO$_2$ monolayer and the Bi2212 substrate
in order to maintain charge neutrality.
We will show that this is indeed supported by density functional theory (DFT) calculations and charge transfer correlations.
Thus, the CuO$_2$ monolayer is highly overdoped and reaches a regime yet inaccessible in bulk cuprates. As shown in Fig.~1, this hole-rich regime approaches 3$d^8$ (Cu$^{3+}$) with $n_e=2$ in the two $e_g$ orbitals. We show that
both \dx and \dz orbitals become active and the electronic structure requires a minimal two-band description with
one electron FS
enclosing $\Gamma$ and one
hole FS around M (Fig.~1). Constructing a two-orbital Hubbard model and studying its SC properties using both weak and strong coupling approaches, we find that the hole-rich CuO$_2$ monolayer is a multiband nodeless superconductor driven by both spin-spin and spin-orbital entangled (super)exchange interactions. The pairing energy gaps are comparable in magnitude to the bulk $d$-wave gap and exhibit a sign-change on the two FS, analogous to Fe-based superconductors. The calculated STM conductance displays the U-shaped spectrum consistent with the experimental observations.

We first carry out a DFT calculation to simulate a CuO$_2$ monolayer on Bi2212 using the Vienna {\em ab initio} simulation package (VASP) \cite{dft1,dft2,dft3,dft4,dft5}. The details are given in the supplemental material (SM) \cite{SM}. As illustrated in Fig.~2(a), due to the missing apical oxygen in the unbalanced octahedron, the cation Cu attracts the bottom anion oxygen (O$_a$) and shortens the out-of-plane Cu-O$_a$ distance at the interface to $2.11$\AA, after relaxation. This value is close to the in-plane Cu-O bond length of $1.92$\AA, which is much shorter than the $2.82$\AA \ Cu-O$_a$ distance in the bulk. The point group symmetry $D_{4h}$ also breaks down to $C_{4v}$. There are two immediate consequences. (i) The $p_z$ orbital of the bottom O$_a$ strongly hybridizes with the Cu \dz orbital. This leads to a \dz like bonding orbital and transfers charges to the oxygen in the BiO layer. The excess charge transfer causes the Cu valence in the monolayer to approch $3d^8$ (Cu$^{3+}$) with two electrons occupying the $e_g$ orbitals. (ii) The crystal field splitting between the \dx and \dz orbitals is significantly reduced compared to in the bulk.

These phenomena show up in the calculated band structure shown in Fig.~2(b). The band highlighted by red markers contains the \dx orbital mixed with the antisymmetric combination of the in-plane oxygen $p_x$ and $p_y$ orbitals
\cite{zhang_rice}. We label this band as the \dx band, which is heavily overdoped and electron-like
near the zone center $\Gamma$. The green markers indicate the \dz band of the Cu \dz orbital mixed with the anion oxygen $p_z$ orbital. It is hole-like near the zone corner $M$, with its band top very close to the Fermi level. Thus, the monolayer CuO$_2$/Bi2212 has a different electronic structure than the CuO$_2$ layer in the bulk. Note that in the experiments \cite{xue}, the CuO$_2$ monolayer is MBE grown on Bi2212 substrates that are optimally hole doped by the excess oxygen dopants of which a substantial fraction resides near the BiO layers \cite{zhoudingwang,ilya}. Their density is further increased in the top BiO layer in the ozone environment. As a result, additional charge transfer takes place via the oxygen dopants hole doping the CuO$_2$ monolayer across the interface, which further stabilizes the CuO$_2$ structure and pushes the chemical potential into the \dz band.
Consequently, the doped holes physically occupy both \dx and \dz orbitals, giving rise to
one electron FS pocket around $\Gamma$ and one hole pocket around $M$ at the Fermi energy. In the cuprates terminology, the CuO$_2$ monolayer corresponds to the heavily overdoped, hole-rich region yet unreachable in bulk materials where the \dz orbital and $d$-$d$ excitations only play a limited role \cite{weber88,cox89,jarrell90,feiner92,zaanen93,feiner96,devereaux10,millis11, aoki10,aoki12}.
\begin{figure}
	\begin{center}
		\fig{3.4in}{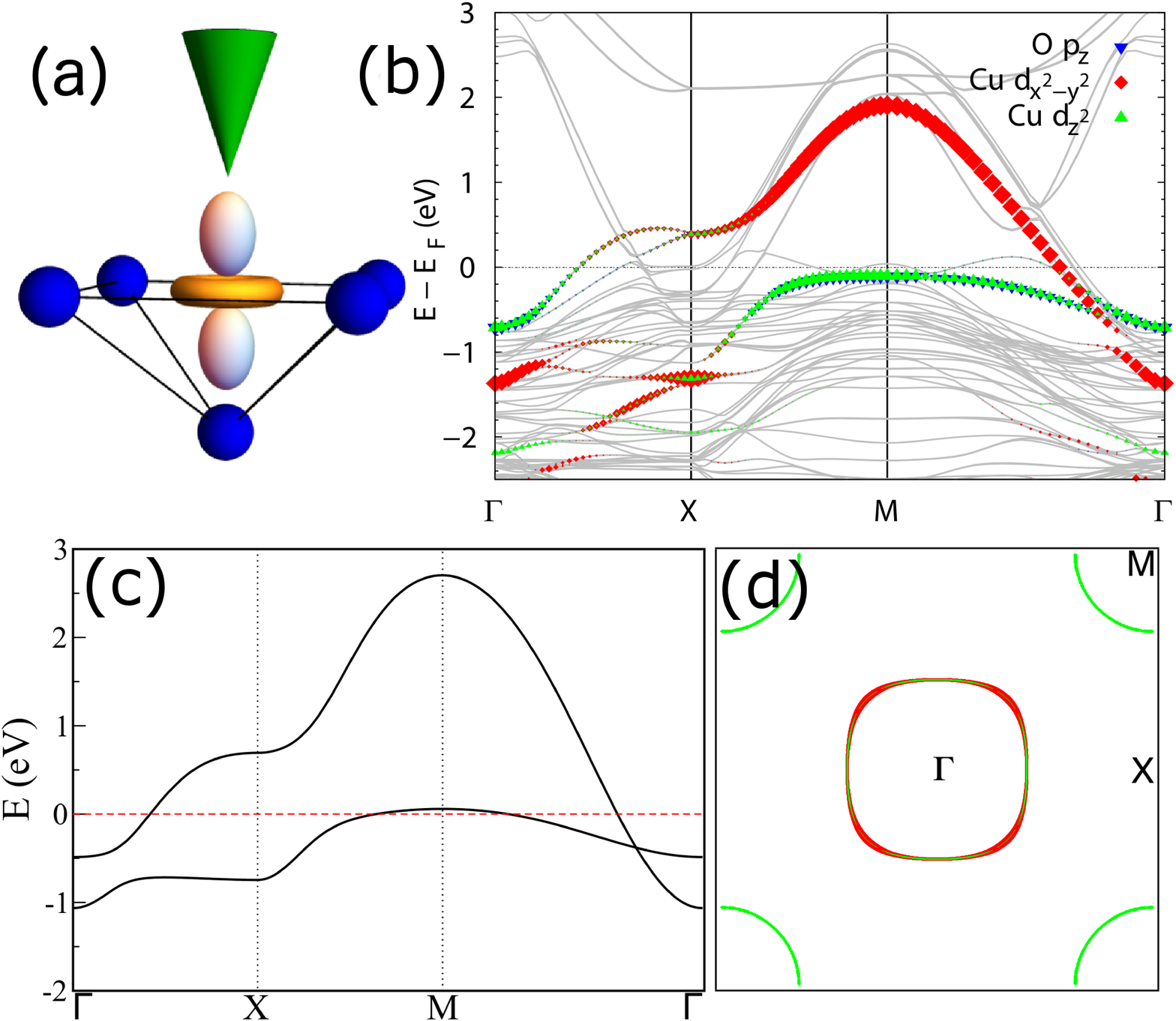}\caption{(a) Atomic structure of monolayer CuO$_2$. Copper \dz orbital is shown in gold and silver (\dx not shown). Blue balls are oxygen ions. The oxygen (O$_a$) at the bottom of pyramid is in the BiO$_2$ layer of Bi2212 substrate.
(b) Band structure of monolayer CuO$_2$/Bi2212 obtained using DFT.
The orbital content of dispersions is colored coded with red ($d_{x^2}$), green ($d_{z^2}$), and blue ($p_z$ strongly hybridized with $d_{z^2}$).
(c) Band dispersion of the two-orbital TB model with FS shown in (d) at $x_h=0.9$
in the same color scheme.
}
	\end{center}
	\vskip-0.5cm
\end{figure}

We next construct a minimal two-orbital Hamiltonian $H=H_t + H_I$ for the monolayer CuO$_2$, where $H_t$ is a tight-binding (TB) model for the band structure and $H_I$ describes the electronic correlations. Using $d_{\alpha\sigma}$, $\alpha=x,z$ to denote a spin-$\sigma$ electron in the \dx and \dz orbitals,
\begin{eqnarray}
H_{t}&=&\sum_{k\alpha\beta\sigma}\varepsilon_k^{\alpha\beta}
d_{k\alpha\sigma}^\dagger d_{k\beta\sigma} + e_z\sum_{k\sigma} d_{kz\sigma}^\dagger d_{kz\sigma},
\label{ht}
\end{eqnarray}
where $e_z$ is the crystal field splitting between the two orbitals. We consider up to third nearest neighbor hopping such that the kinetic energy of intraorbital hopping in Eq.~(\ref{ht}) is $\varepsilon_k^{\alpha\alpha}=-2t_\alpha \gamma_k-4t_\alpha^\prime \alpha_k-2t_\alpha^{\prime\prime}\gamma^\prime_k$ with lattice harmonics of $A_1$ symmetry $\gamma_k=\cos k_x+\cos k_y$, $\alpha_k=\cos k_x\cos k_y$, and $\gamma_k^\prime=\cos2k_x+\cos2k_y$. Due to the different orbital symmetry, the interorbital hopping leads to $\varepsilon_k^{xz}=2t_{xz} \beta_k+2t_{xz}^{\prime\prime}\beta^\prime_k$, with $B_1$ harmonics $\beta_k=\cos k_x-\cos k_y$ and $\beta_k^{\prime}=\cos2 k_x-\cos 2k_y$.
The parameters of the TB model are given in the SM \cite{SM} and the chemical potential is treated as an independent variable.
The TB band structure is shown in Fig.~2(c) at doping $x_h=0.9$ or $n_e=2.1$. It describes the DFT results in Fig.~2(b) very well. The FS is plotted in Fig.~2(d), showing
one electron pocket around $\Gamma$ and one hole pocket around M. Since $\varepsilon_k^{xz}$ has $d$-wave form factors,
the FS around $\Gamma$ is mostly \dx like around nodal but of a mixed character around antinodal directions.
The hole pocket around M is mainly \dz like since the bands are well separated in energy. The smaller overlap of out-of-plane orbitals makes the \dz band narrow with a small bandwidth, consistent with the DFT dispersions.

The correlation part follows from the standard two-orbital Hubbard model \cite{castellani,georges} for the $e_g$ complex
\begin{align}
H_I&=U\sum_{i,\alpha}\hn_{i\alpha\upa}\hn_{i\alpha\dna}
+\left(U'-{1\over 2}J_H\right)\sum_{i,\alpha<\beta}\hn_{i\alpha}\hn_{i\beta}
\label{hi} \\
&-J_H\sum_{i,\alpha\neq\beta}{\bf S}_{i\alpha}\cdot {\bf S}_{i\beta}
+J_H\sum_{i,\alpha\neq\beta}d^\dg_{i\alpha\upa}
d^\dg_{i\alpha\dna}d_{i\beta\dna}d_{i\beta\upa},
\nonumber
\end{align}
where the intra and interorbital Coulomb $U$ and $U^\prime$ are related to Hund's coupling $J_H$ by $U=U'+2J_H$.

The emergence the low-energy $d_{z^2}$-band and the hole FS pocket around M enables an analogy to multiorbital nodeless Fe-pnictides superconductors \cite{mazin,hirshfeld-scalapino,frg,
chubukov08,hu,fczhang,kontani,yanagi,zhoukotliarwang}, particularly when the correlation effects in Eq.~(2) are treated using weak-coupling approaches. Since perfect nesting between the electron and hole pockets is absent,
the only logarithmic divergence is in the Cooper channel. Following Ref.~\cite{chubukov08}, we performed a two-patch renormalization group analysis and found that the nodeless $s$-wave superconductivity is the leading instability. Moreover, the relevant pair-scattering across the electron and hole FS, i.e. the $u_3$ channel \cite{chubukov08}, drives a sign-changing $s_\pm$ gap function on the two FS pockets.

An important difference between the Fe-based and cuprate superconductors is, however, that the former are $p$-$d$ charge-transfer metals, whereas the latter are charge-transfer insulators \cite{zhoukotliarwang}. Indeed, the nodeless SC state of monolayer CuO$_2$/Bi2212 emerges \cite{xue} inside a charge-transfer gap of similar magnitude as in bulk cuprates \cite{yywang}. It is thus necessary to carry out a strong coupling study of the two-orbital Hubbard model. To this end, we derive in the SM the general spin-orbital superexchange interactions of the Kugel-Khomskii type \cite{SM,kk,castellani},
\begin{eqnarray}
H_{\rm J-K}=\sum_{\langle ij\rangle}\biggl[ J{\bf S}_{i}\cdot{\bf S}_j
&+&\sum_{\mu\nu} I_{\mu\nu} T_i^\mu T_j^\nu
\label{hs} \\
&+&\sum_{\mu\nu} K_{\mu\nu}({\bf S}_{i}\cdot{\bf S}_j)(T_i^\mu T_j^\nu)\biggr]
\nonumber
\end{eqnarray}
where ${\bf S}_i$ is the spin-1/2 operator, $T_i^\mu$, $\mu=0,x,y,z$, are the orbital pseudospin-1/2 operators in the orbital basis $(\vert x^2-y^2\rangle,\vert z^2\rangle)^T$ \cite{kk}.
In Eq.~(\ref{hs}), the $J$-term is the SU(2) invariant Heisenberg spin exchange coupling, while the terms proportional $I_{\mu\nu}$ and $K_{\mu\nu}$
describe the anisotropic orbital and spin-orbital entangled superexchange interactions respectively, since the orbital/pseudospin rotation symmetry is broken by the generic hoppings and crystal field in $H_t$.
We thus arrive at an effective two-orbital strong coupling model
\begin{equation}
H=P_GH_tP_G+ H_{\rm J-K},
\label{htjk}
\end{equation}
where $P_G$ stands for the Gutzwiller projection of states with multiple occupations. Hereafter, we consider
Eq.~(\ref{htjk}) as an effective low-energy theory for the hole-rich regime of monolayer CuO$_2$ and study the emergent SC state due to the spin-orbit superexchange correlations.
The Gutzwiller projection is treated in the SM \cite{SM} using the variational Gutzwiller approximation \cite{gebhard98,lechermann,sen} for a generic set of interactions $U=2.5$eV and $J_H=0.1U$.

The intersite quantum spin-orbital fluctuations described by Eq.~(\ref{hs}) can be projected into the spin-singlet channel by $P_{ij}^s={\bf S}_{i}\cdot{\bf S}_j-1/4$, and written in terms of the pairing operators $\Delta_{ij}^{\alpha\beta\dagger}=d_{i\alpha\uparrow}^\dagger d_{j\beta\downarrow}^\dagger-d_{i\alpha\downarrow}^\dagger  d_{j\beta\uparrow}^\dagger$.
Since the \dx and \dz orbitals are split by the crystal field, there is an orbital order that causes the operator $T_{iz}$ in Eq.~(\ref{hs}) to take on its expectation value $1/2$. As shown in the SM \cite{SM}, this leads to a spin exchange interaction corresponding to that of Heisenberg term in the $t$-$J$ model \cite{lee} with the familiar result $J_s({\bf S}_{i}\cdot{\bf S}_j-{1\over4}n_in_j)=-{J_s\over2} \sum_{\alpha\beta}\Delta_{ij}^{\alpha\beta\dagger} \Delta_{ij}^{\alpha\beta}$.
We set $J_s=120$meV, the commonly accepted value for bulk cuprates \cite{lee}. However, the $T_{iz}$ order does not quench the transverse orbital fluctuations represented by $T_i^{\pm}$ that contribute to pairing. Remarkably, such spin-orbit entangled, quadruple exchange interactions in Eq.~(\ref{hs}) generates a new pairing contribution
\begin{equation}
K P_{ij}^s(T_i^+T_j^+ + h.c.)
= - {K\over 2}(\Delta_{ij}^{xx\dagger}\Delta_{ij}^{zz} + h.c.)
\label{kpp}
\end{equation}
which captures the physics of the interorbital pair scattering. This is the strong coupling counterpart of the inter FS pocket pair scattering in weak-coupling approaches \cite{cox89,chubukov08}. We considered all spin-singlet pairing in the SM \cite{SM} and determined the expectation values of the pairing fields $\langle \Delta_{ij}^{\alpha\beta}\rangle$ self-consistently in the Gutzwiller approximation. The latter has the form,
\begin{eqnarray}
\langle \Delta_{ij}^{\alpha\beta}\rangle&=&\frac{1}{N_s}
\sum_{\mathbf{k},\alpha\beta}\Delta_{\alpha\beta}
b_{\alpha\beta}(\mathbf{k})e^{i\mathbf{k}(r_i-r_j)},
\label{pairing}
\end{eqnarray}
where $N_s$ is the number of lattice sites and $b_{\alpha\beta}(k)$ the form factors of different symmetries in the $C_{4v}$ point group of the crystal. For nearest neighbor pairing, $b_{\alpha\alpha}(k)=\gamma_k$ and $b_{xz}(k)=\beta_k$ in the $A_1$ symmetry channel whereas $b_{\alpha\alpha}(k)=\beta_k$ and $b_{xz}(k)=\gamma_k$ in the $B_1$ channel.
Our results show that the variational ground state in the strong coupling theory is a nodeless superconductor with $A_1$ symmetry in the hole-rich regime where the FS contains both the electron and hole pockets. Moreover, the pairing fields are dominated by the intra-orbital $\Delta_{\alpha\alpha}$ with extended $s$-wave form factor $b_{\alpha\alpha}=\gamma_k$ in Eq.~(\ref{pairing}).
In Figs.~3(a-b), we show the FS at $x_h=0.9$ and the pairing energy gaps as a function of the angle along the two FS pockets for $K=80$meV. The nodeless $s_\pm$ gap function with opposite signs and
comparable magnitude is a remarkable consequence of the spin-orbital entangled exchange pairing interaction $K$ in Eq.~(\ref{kpp}). The momentum space anisotropy of the gap function is small and more apparent on the electron pocket around $\Gamma$ which is larger and less circular.
Such a nodeless multiorbital superconductor in the hole-rich regime is proposed in Fig.~1 for the monolayer CuO$_2$/Bi2212.
\begin{figure}
	\begin{center}
		\fig{3.4in}{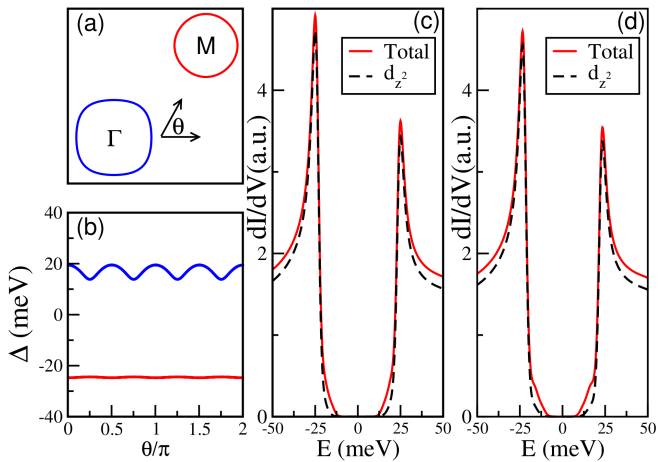}\caption{(a) Normal state FS at $x_h=0.9$.
(b) SC pairing energy gaps at $K=80$meV along FS pockets with angle $\theta$ defined in (a). (c) U-shaped total LDOS (red solid line) showing a nodeless gap $\sim25$meV. LDOS from \dz orbital is shown in dashed lines. (d) Same as in (c) obtained for $K=60$meV.}
	\end{center}
	\vskip-0.5cm
	\label{scorder}
\end{figure}

To compare to STM, we calculate the LDOS for each orbital $N_\alpha(\omega)=\sum_{k\sigma}{\rm Im}\int_0^\beta e^{{i\omega}\tau}\langle{\rm T}_\tau d_{k\alpha\sigma}(\tau)d_{k\alpha\sigma}^\dagger(0)\rangle$. The total LDOS,
$N(\omega)=N_x(\omega)+N_z(\omega)$,
is shown in Fig.~3(c). It has a U-shaped spectrum with a pair of coherence peaks demarcating a nodeless energy gap around $25$meV, in good agreement with STM observations \cite{xue}. The LDOS from the \dz orbital, also shown in Fig.~3(c), has a slightly larger onset spectral gap and the majority of the
total differential conductance due to the large DOS of the \dz band.
Fig.~3(d) shows the LDOS spectra obtained for a smaller $K=60$meV. The smaller onset spectral gap associated with the \dx orbital is visible, but the tunneling conductance, especially that into the \dz orbital, continues to exhibit the U-shaped spectrum. The tunneling matrix element on top of the monolayer also favors a path through the out-of-plane Cu 3\dz orbital as depicted in Fig.~2(a).
The overall agreement with the STM findings supports our conjecture that the electronic structure and correlation in the monolayer
CuO$_2$/Bi2212 produce an intrinsic two-orbital nodeless $s$-wave SC state.

The proposed two-orbital nodeless SC state near Cu $3d^8$ is different from the single-orbital extended $s$-wave pairing state known to arise with very small pairing amplitude in the overdoped single-band $t$-$J$ model \cite{ubbens-lee,lee}. To verify this point, we studied the intermediate doping regime
$0.3<x_h<0.7$, where the \dz hole pocket has disappeared following a Lifshitz transition and the FS contains a single \dx electron pocket enclosing $\Gamma$ (Fig.~S1(c) in SM). The SC state indeed has extended $s$-wave pairing amplitudes that are two orders of magnitude smaller due to the suppression of orbital fluctuations. The electron pocket grows with reducing doping and transitions to a large hole FS around $M$ for $x_h<0.3$, where the nodal $d$-wave SC state is recovered as in the bulk cuprates. Finally, when the two-orbital model is studied at $x_h=1$, i.e. in the Cu $3d^8$ limit with two electrons in the $e_g$ complex, the Gutzwiller projected Hamiltonian $P_GH_tP_G$ has an insulating ground state with AF long-range order for our parameters as indicated in the phase diagram Fig.~1.
This is consistent with the high-spin Mott insulating state of the two-orbital Hubbard model at half-filling
\cite{ruegg05,biermann,millis07,werner16}, where the AF spin moments from the two orbitals are ferromagnetically aligned by $J_H$ \cite{hqlin12}.

In conclusion, we proposed that the ozone MBE grown CuO$_2$ monolayer on Bi2212 is heavily overdoped due to interface charge transfer, reaching the hole-rich regime yet inaccessible in bulk cuprates. The resulting electronic structure involves
holes occupying both Cu 3\dx and 3\dz orbitals. The quantum fluctuations of the spin-orbital superexchange interaction are shown to produce a two-orbital nodeless superconductor with a U-shaped LDOS and a comparably sized pairing gap as in bulk Bi2212 near optimal doping, providing a natural explanation of the STM experiments \cite{xue}. Although the SC proximity effect between a $d$-wave cuprate and a normal metal is difficult to achieve in $c$-axis oriented junctions \cite{proximity} and more detailed studies are necessary, it is reasonable to expect that the intrinsic nodeless SC state of the CuO$_2$ monolayer to establish phase coherence with the bulk $d$-wave superconductor through inhomogeneous Josephson coupling at the interface.
A possible mechanism to facilitate the interface charge transfer is through the type-B oxygen dopants in Bi2212 \cite{zhoudingwang}, residing close to the BiO layer as observed by STM \cite{ilya}. Ozone MBE growth can increase significantly the type-B dopants on the surface BiO layer, which in turn provide heavy hole-doping for the capping monolayer CuO$_2$. The predictions can be tested experimentally by measuring the quasiparticle band dispersion using ARPES or STM quasiparticle interference on samples with large enough coverage of high quality CuO$_2$ monolayer. Indeed, interface charge transfer and the change of FS topology have been observed recently with enhanced $T_c$ in monolayer FeSe superconductors grown on SrTiO$_3$ substrate \cite{xue12,xjzhou12,xjzhou13,ma14}. It would also be interesting to probe and study the phonon dynamics and electron-phonon coupling at the interface \cite{phonon}.
The findings presented here provide insights for a new direction of searching for high-T$_c$ superconductors in extended doping regimes and with liberated orbital degrees of freedom in ozone MBE grown transition metal oxides heterostructures.

We thank Sen Zhou and Andy Millis for helpful discussions. This work is supported in part by the Ministry of Science and Technology of China 973 program (No.~2017YFA0303100, No. 2015CB921300), National Science Foundation of China (Grant No. NSFC-1190020, 11534014, 11334012), and the Strategic Priority Research Program of CAS (Grant No.XDB07000000); and the U.S. Department of Energy, Basic Energy Sciences Grant No. DE-FG02-99ER45747 (K.J. and Z.W.). Z.W. thanks the hospitality of IOP, CAS and Aspen Center for Physics and the support of ACP NSF grant PHY-1066293.

\newpage
\renewcommand{\theequation}{S\arabic{equation}}
\renewcommand{\thefigure}{S\arabic{figure}}
\renewcommand{\thetable}{S\arabic{table}}
\setcounter{equation}{0}
\setcounter{figure}{0}

\title{Supplemental Material: Nodeless high-T$_c$ superconductivity in highly-overdoped monolayer CuO$_2$}

\maketitle
\section*{Supplemental Material}

\subsection{Density functional calculations}

We performed density functional theory (DFT) calculations employing the projector augmented wave (PAW) method encoded in the Vienna ab initio simulation package (VASP) \cite{SKresse1993,SKresse1996,SKresse1996b}. Generalized-gradient approximation (GGA) \cite{SPerdew1996} for the exchange correlation functional is used. The cutoff energy is set to be 500 eV for expanding the wave functions in the plane-wave basis.  The Brillouin zone is sampled in $\textbf{k}$ space within Monkhorst-Pcak scheme \cite{SMonkhorst1976} and the number of $\textit{k}$ points depends on the lattice: ${11}\times{ 11 }\times{ 1}$.
The surface is modeled using a periodically repeated slab consisting of eight CuO$_2$ layers (including the surface CuO$_2$ layers) and six BiO layers plus a vacuum layer of 20 \AA~ with inversion symmetry through the center of the
slab. The inner CuO$_2$ and BiO layers are frozen while the surface CuO$_2$ and BiO layers are allowed to relax with forces minimized to less than 0.02 eV/\AA. The results of the DFT calculations are shown in Fig.~2(b).

The formation energy of the adopted monolayer structure in the DFT calculation is about -2.2 eV per CuO$_2$, thus yielding a stable structure. It is instructive to compare this to the alternative CuO structure \cite{Sfawang}. In Ref.9 of the main text, the formation energy for Cu$_2$O$_2$ is only about -0.78 eV lower than that of CuO$_2$ (Note that there are two Cu$_2$O$_2$ formulas in that calculation). Thus, the formation energies for the two stable structures are comparable.
Which stable structure is realized in the experiments depends on both the formation energy and the experimental chemical environments. In some cases, structures with less negative formation energies can be realized. In the main text, we pointed out that the oxygen dopants in the close to optimally doped Bi2212 substrates and their increased density in the top BiO layer in the ozone growth environment may promote additional dopant-induced charge transfer across the interface and play an important part in realizing the CuO$_2$ monolayer. Thus, the DFT calculations of the ideal structures cannot determine uniquely the structure at the present time and we have to rely on experimental findings. While there is no evidence that would support the formation of the tetragonal CuO structure, there are indeed many pieces of experimental evidence supporting the CuO$_2$ structure in the monolayer. (1) In the STM measurements (Ref.5 in the main text), the distance between nearest neighbor (NN) Cu ions is about 3.8 \AA, which is consistent with the CuO$_2$ structure but inconsistent with CuO where the distance between NN Cu ions is about 2.69 \AA. (2) The STM spectra show that the separation between the valence band and the Fermi level is close to 1.40 eV, which is much smaller than that expected of the tetragonal CuO layer (about 2.35 eV). (3) The experimentally determined distance between the copper-oxygen monolayer and the top BiO layer in the Bi2212 substrate is about 1.73 \AA~[Fig.1(a) in Ref.5], which is in very good agreement with the 1.67 \AA~in the calculations using the CuO$_2$ structure. If the surface layer were to crystalize in the CuO structure, this separation would be about 2.9 \AA~according to our DFT calculations. This gives additional support for the monolayer to crystalize in the stable CuO$_2$ structure in the experiment.
\begin{figure}
	\begin{center}
		\fig{3.4in}{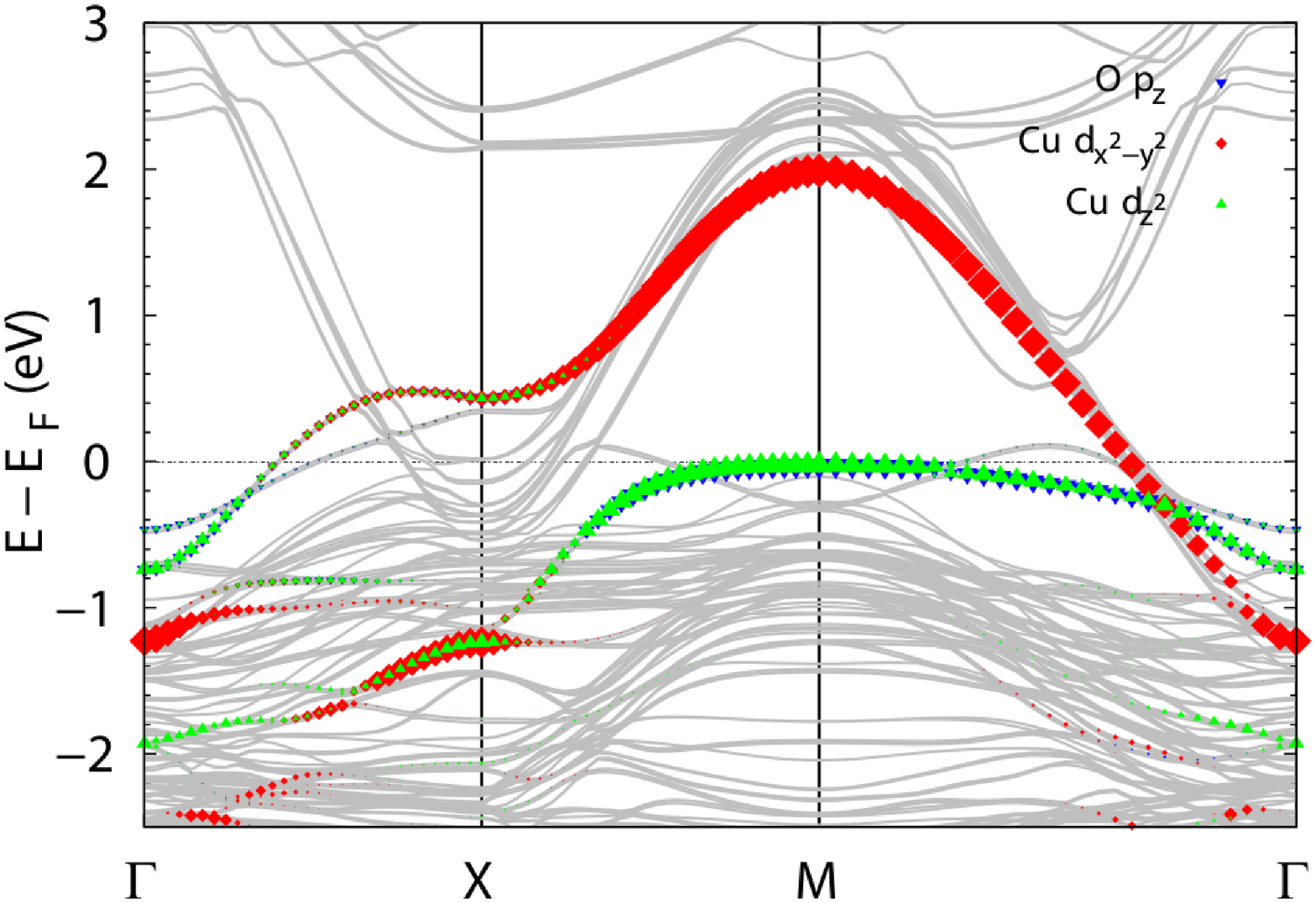}\caption{
			Band structure of monolayer CuO$_2$/Bi2212 obtained with slabs containing 10 bulk CuO$_2$ layers.
			The orbital content of dispersions is colored coded with red ($d_{x^2}$), green ($d_{z^2}$), and blue ($p_z$ strongly hybridized with $d_{z^2}$).
		}
	\end{center}
	\vskip-0.5cm
\end{figure}

It is important to study the stability of the electronic band structure and the atomic structural parameters
obtained for the CuO$_2$ monolayer under changing slab thickness in the DFT calculations. In Fig.~S1, the band structure obtained for slabs with 10 bulk CuO$_2$ layers is shown, which is very close to the band structure obtained with 8 bulk CuO$_2$ layers shown in Fig.~2b in the main text. Moreover, the structural parameters at the interface between the surface CuO$_2$ and the top BiO layer exhibit only small variations for slabs of different thickness. In Table S1, the values of the CuO$_2$ surface layer buckling, the bond length of the Cu and the anion oxygen O$_a$, as well as the distance between the CuO$_2$ monolayer and the top BiO layer (as determined by the vertical distance between the O and Bi atoms) are listed for comparison. Thus, we conclude that the obtained band structure and the structural parameters are stable to the changes in the slab thickness in the DFT calculations.
\begin{table}[t]
	\caption{Structural parameters for different slab thickness}
	
	\begin{tabular}{c|c|c}
		\hline
		\hline
		& 8 CuO$_2$ layers & 10 CuO$_2$ layers \\	
		\hline
		CuO$_2$ layer buckling & 0.199\AA & 0.212\AA  \\	
		\hline
		Cu-O$_a$ bond length & 2.115\AA & 2.112\AA \\	
		\hline
		CuO$_2$-BiO distance & 1.452\AA & 1.461\AA \\	\hline \hline
	\end{tabular}
\end{table}

\subsection{Tight-binding model parameters and doping evolution of the Fermi Surface}

Based on the DFT results, we construct a two-orbital tight-binding (TB) model of Cu $e_g$ complex for the monolayer CuO$_2$. The Hamiltonian is given in Eq.~(1) in the main text. Denoting $d_{\alpha\sigma}$, $\alpha=x$($d_{x^2}$)$,z$($d_{z^2}$)
\begin{eqnarray}
	H_{t}&=&\sum_{k\sigma}\varepsilon_k^{xx}
	d_{kx\sigma}^\dagger d_{kx\sigma}
	+ \sum_{k\sigma}\varepsilon_k^{xz}
	(d_{kx\sigma}^\dagger d_{kz\sigma}+h.c.)
	\nonumber \\
	&+&\sum_{k\sigma}\varepsilon_k^{zz}
	d_{kz\sigma}^\dagger d_{kz\sigma}+ e_z\sum_{k\sigma} d_{kz\sigma}^\dagger d_{kz\sigma}
	\label{Sht}
\end{eqnarray}
where $\varepsilon_k^{\alpha\beta}$ is the kinetic energy due to intra and interorbital hopping, and $e_z$ is the crystal field splitting between \dz and \dx orbitals. Up to third nearest neighbor hopping, we have
\begin{eqnarray}
	\varepsilon_k^{xx}&=&-2t_x\gamma_k-4t_x^\prime \alpha_k-2t_x^{\prime\prime}\gamma^\prime_k
	\nonumber \\
	\varepsilon_k^{zz}&=&-2t_z \gamma_k-4t_z^\prime \alpha_k-2t_z^{\prime\prime}\gamma^\prime_k
	\nonumber \\
	\varepsilon_k^{xz}&=&2t_{xz} \beta_k+2t_{xz}^{\prime\prime}\beta^\prime_k
	\label{Sek}
\end{eqnarray}
where the intraorbital hopping involves lattice harmonics of $A_1$ symmetry $\gamma_k=\cos k_x+\cos k_y$, $\alpha_k=\cos k_x\cos k_y$, and $\gamma_k^\prime=\cos2k_x+\cos2k_y$, and the interorbital hopping involves $B_1$ harmonics $\beta_k=\cos k_x-\cos k_y$ and $\beta_k^{\prime}=\cos2 k_x-\cos 2k_y$.
The hopping parameters for the 1st ($t$), 2nd ($t^\prime$), and 3rd ($t^{\prime\prime}$) nearest neighbors are listed in Table S2. The crystal field splitting between \dx and \dz  is $e_z=-0.91$eV. This set of band parameters and the band structure of the two orbital model is similar to the bulk electronic structure of the single-layer cuprates La$_{2}$(Sr/Ba)CuO$_4$ \cite{Saoki10,Saoki12}. However, the results obtained in this paper do not dependent on the detailed choices of the band parameters.

In Fig.~S2, we show the evolution of the Fermi surface (FS) in the two-orbital TB model as a function of the hole doping concentration $x_h$. The corresponding electron concentration in the $e_g$ orbitals is given by $n_e=3-x_h$.
There are three different types FS topology, thus three doping regimes, between $x_h=0$ ($d^9$) and $x_h=1.0$ ($d^8$). In regime I shown Figs.~S2(a-b), $x_h$ is small; there is only one large hole FS pocket enclosing $M(\pi,\pi)$. This is the typical region in bulk cuprates below moderate level of overdoping. With increasing hole-doping, the hole pocket changes via a Lifshitz transition to an electron pocket enclosing $\Gamma$ near $x_h=0.3$ and $n_e\approx2.7$. The FS in this regime II is shown Fig.~S2(c), which may still be realizable in very overdoped bulk cuprates. Further increasing doping leads to the hole-rich regime III  via a second Lifshitz transition associated with the emergence of a new hole FS pocket around $M$ near $x_h\approx0.7$ and $n_e\approx2.3$. Figs.~S2(d-f) exemplify the FS in regime III. This  hole-rich region is proposed here for the ozone MBE grown CuO$_2$ monolayer on top of Bi2212; it has both electron and hole FS pockets and is yet unreachable in bulk cuprates.
\begin{table}[t]
	\caption{Hopping parameters of the TB model in eV.}
	
	\begin{tabular}{c|c|c|c}
		\hline
		\hline
		hopping integral & 1rd ($t$) & 2nd ($t^\prime$) & 3rd ($t^{\prime\prime}$)\\	\hline
		intra-orbital $t_x$ & 0.471 & -0.0932 & 0.0734 \\	\hline
		intra-orbital $t_z$ & 0.0682 & 0.0109 & 0.0\\	\hline
		inter-orbital $t_{xz}$ & 0.178 & 0.0 & 0.0258 \\	\hline \hline
	\end{tabular}
\end{table}
\begin{widetext}
	
	\begin{figure}
		\begin{center}
			\fig{5.0in}{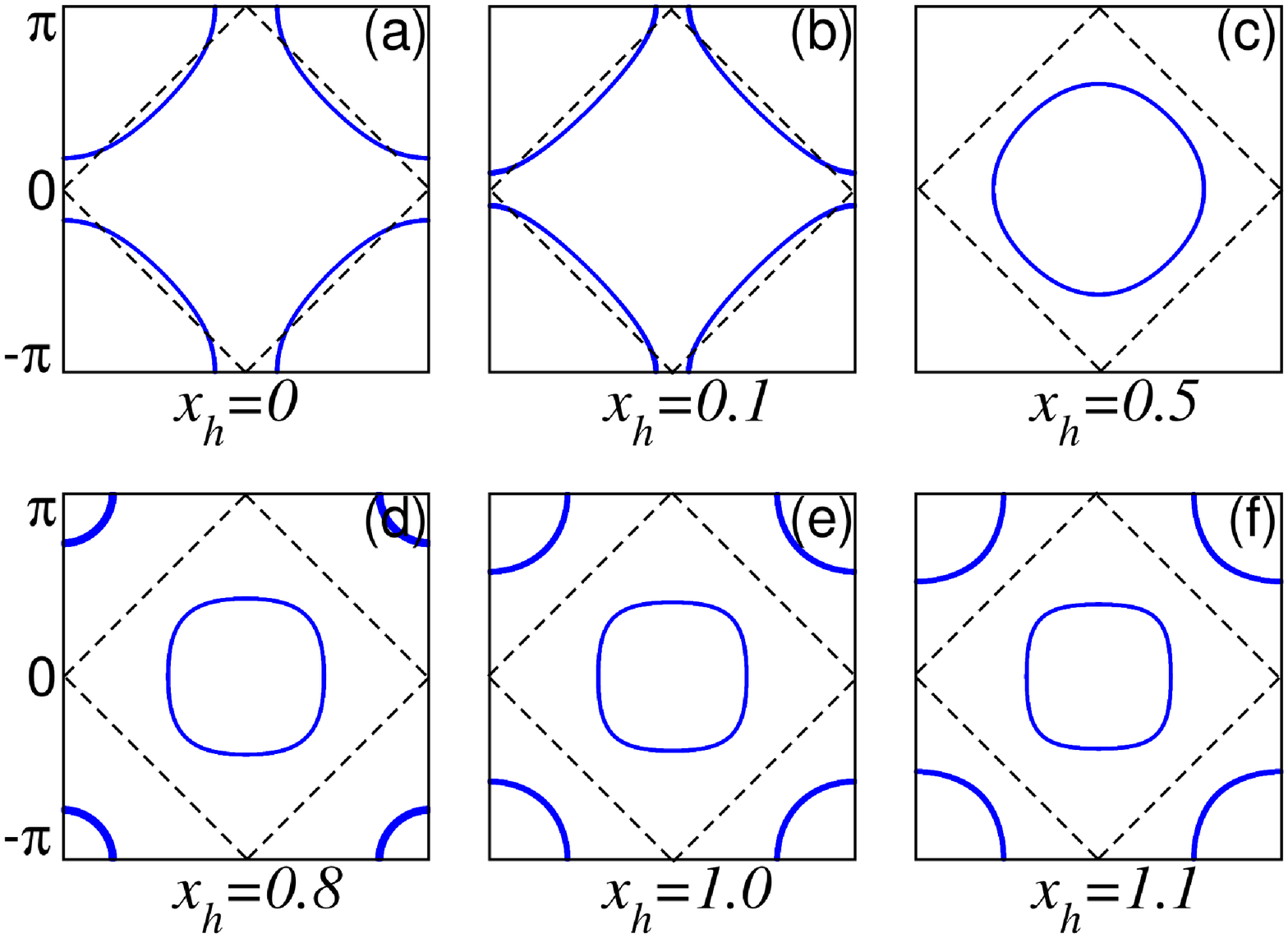}\caption{FS evolution as a function of the hole doping concentration $x_h$ showing three different regimes with different FS topology. Regime I(a-b): There is one hole pocket around $M$. Regime II(c): There is only one electron pocket around $\Gamma$. Regime III (e-f): there are one electron pocket around $\Gamma$ and one hole pocket around $M$. Dashed lines are nodal lines for the extend $s$-wave pairing.}
		\end{center}
		\vskip-0.5cm
	\end{figure}
	
	\subsection{Multi-orbital Gutzwiller approximation}
	
	To treat multi-orbital Hubbard interactions in the strong coupling regime appropriately, we apply the multi-orbital Gutzwiller approach \cite{Sgebhard98,Ssen}, which is equivalent to the slave boson formalism \cite{SLechermann}.
	The standard multi-orbital Hubbard correlations are described by
	\begin{eqnarray}
		\hH_I=\sum_i\hH_{Ii}=U\sum_{i,\alpha}\hn_{i\alpha\upa}\hn_{i\alpha\dna}
		+\left(U'-{1\over 2}J_H\right)\sum_{i,\alpha<\beta}\hn_{i\alpha}\hn_{i\beta}
		\label{h1} -J_H\sum_{i,\alpha\neq\beta}{\bf S}_{i\alpha}\cdot {\bf S}_{i\beta}
		+J_H\sum_{i,\alpha\neq\beta}d^\dg_{i\alpha\upa}
		d^\dg_{i\alpha\dna}d_{i\beta\dna}d_{i\beta\upa}, \label{hub}
	\end{eqnarray}
	with $U=U'+2J_H$. The local interaction $\hH_{Ii}$ at each site $i$ can be written in the basis of the local Fock states,
	\begin{eqnarray}
		\hH_{Ii}=\sum_{n,n'}|i,n\rangle H_{i}^{nn'}\langle i,n'|,\quad H_{i}^{nn'}=\langle i,n|\hH_{Ii}|i,n'\rangle,
	\end{eqnarray}
	in the occupation representation,
	\begin{eqnarray}
		|i,n\rangle=|{n_{i\alpha\sigma}}\rangle=\prod_{\alpha\sigma}
		(c_{i\alpha\sigma}^\dagger)^{n_{i\alpha\sigma}}|vac\rangle,
		\nonumber
	\end{eqnarray}
	where $|vac\rangle$ is the vacuum state and $n_{i\alpha\sigma}=0,1$ is the local occupation.
	$\hH_{Ii}$ can now be diagonalized to obtain
	\begin{eqnarray}
		\hH_{Ii}=\sum_{\Gamma}E_{\Gamma}|i,\Gamma\rangle\langle i,\Gamma|
	\end{eqnarray}
	where the eigen-Fock state $|i,\Gamma\rangle$ is a unitary transformation of the Fock stae $|i,n\rangle$, $|\Gamma\rangle=\sum_{n}T_{\Gamma n}|i,n\rangle$. The eigen-state energy $E_{\Gamma}$ and
	the transformation matrix $T_{\Gamma n}$ are site-independent since $\hH_{Ii}$ is translation invariant. In view of the slave-boson formulation, one can associate an auxiliary boson $\phi_{i\Gamma}$ with each eigen-Fock state $|i,\Gamma\rangle$. The expectation $\langle\phi_{i\Gamma}^{\dagger}\phi_{i\Gamma}\rangle$ measures the probability of occupying the eigen-Fock state $|i,\Gamma\rangle$.
	
	The physical electron operator can therefore be represented by
	\begin{eqnarray}
		d_{i\alpha\sigma}^\dagger&=&z_{i\alpha\sigma}^\dagger f_{i\alpha\sigma}^\dagger \\
		z_{i\alpha\sigma}&=&L_{i\alpha\sigma}R_{i\alpha\sigma}\sum_{\Gamma,\Gamma'}D_{\Gamma\Gamma'}^{\alpha\sigma}\phi_{i\Gamma}^{\dagger}\phi_{i\Gamma'}
	\end{eqnarray}
	with local constraints
	\begin{eqnarray}
		\sum_{\Gamma}\phi_{i\Gamma}^{\dagger}\phi_{i\Gamma}&=&1 \\
		\sum_{\Gamma\Gamma'}n_{\alpha\sigma}^{\Gamma\Gamma'}\phi_{i\Gamma}^{\dagger}\phi_{i\Gamma'}&=&f_{i\alpha\sigma}^\dagger f_{i\alpha\sigma}, \qquad n_{\alpha\sigma}^{\Gamma\Gamma'}=\langle \Gamma| \hat{n}_{\alpha\sigma} |\Gamma' \rangle.
	\end{eqnarray}
	At the mean-field level, which is equivalent to the Gutzwiller approximation, all the bosons are condensed to their expectation values with $\phi_{i\Gamma}^\dagger=\phi_{i\Gamma}$, $L_{i\Gamma\sigma}=\frac{1}{\sqrt{n_{i\alpha\sigma}}}$, $R_{i\Gamma\sigma}=\frac{1}{\sqrt{1-n_{i\alpha\sigma}}}$, and
	\begin{eqnarray}
		D_{\Gamma\Gamma'}^{\alpha\sigma}&=&\langle \Gamma| d_{\alpha\sigma}^\dagger |\Gamma' \rangle=\sum_{n,n'}T_{\Gamma n}T_{\Gamma' n'}\langle n| d_{\alpha\sigma}^\dagger|n'\rangle.
	\end{eqnarray}
	The Hubbard Hamiltonian in the Gutzwiller approximation thus becomes
	\begin{eqnarray}
		P_GH_tP_G&=&\sum_{i\alpha,j\beta,\sigma}[z_{i\alpha\sigma}z_{j\beta\sigma}t_{i\alpha,j\beta}+(\epsilon_{i\alpha\sigma}+e_{\alpha})\delta_{ij}\delta_{\alpha\beta}]f_{i\alpha\sigma}^\dagger f_{j\beta\sigma}
		+\sum_{i,\Gamma}E_{\Gamma}\phi_{i\Gamma}^2+\sum_{i}\lambda_{i}(\sum_\Gamma\phi_{i\Gamma}^2-1)-\sum_{i\alpha\sigma}\epsilon_{i\alpha\sigma}n_{i\alpha\sigma}   \nonumber
	\end{eqnarray}
	with $ z_{i\alpha\sigma}=\frac{1}{\sqrt{n_{i\alpha\sigma}(1-n_{i\alpha\sigma})}}
	\sum_{\Gamma\Gamma'}D_{\Gamma\Gamma'}^{\alpha\sigma}\phi_{i\Gamma}
	\phi_{i\Gamma'}
	$.
	The hopping integral $t_{i\alpha,j\beta}$ is seen to be renormalized by $z_{i\alpha\sigma}z_{j\beta\sigma}$, whereas $e_{\alpha}$ is the effective local crystal field, and $\lambda_{i}$ and $\epsilon_{i\alpha\sigma}$ are the Lagrange multipliers associated with the constraints.
	
	To study superconductivity using this strong coupling approach, we include the Kugel-Khomskii spin-orbital superexchange interactions described by $H_{\rm J-K}$ derived in the following section D. The effective mean-field Hamiltonian split into fermionic and bosonic parts. The fermionic part is given by
	\begin{eqnarray}
		\hH_f=\sum_{i\alpha,j\beta,\sigma}[z_{i\alpha\sigma}z_{j\beta\sigma}t_{i\alpha,j\beta}+(\epsilon_{i\alpha\sigma}+e_{\alpha})\delta_{ij}\delta_{\alpha\beta}]f_{i\alpha\sigma}^\dagger f_{j\beta\sigma}+\hH_{\rm J-K}^\Delta
		\label{h-gutzwiller}
	\end{eqnarray}
	where $\hH_{\rm J-K}^\Delta$ is the decoupled spin-orbital exchange interaction in the pairing channel discussed in detail in section D and is quadratic in the $f$ operators. Thus the fermionic part can be diagonalized by solving the BdG equations to obtain the expectation values
	\begin{eqnarray}
		\langle n_{i\alpha\sigma} \rangle &=& \langle f_{i\alpha\sigma}^\dagger f_{i\alpha\sigma} \rangle \nonumber \\
		\langle  \Delta_{ij}^{\alpha\beta\dagger} \rangle &=&\langle  f_{i\alpha\uparrow}^\dagger f_{j\beta\downarrow}^\dagger-f_{i\alpha\downarrow}^\dagger  f_{j\beta\uparrow}^\dagger \rangle \nonumber \\ \langle \Delta_{ij}^{\alpha\beta} \rangle &=&\langle  f_{j\beta\downarrow}f_{i\alpha\uparrow}-f_{j\beta\uparrow}f_{i\alpha\downarrow} \rangle.
		\label{pp}
	\end{eqnarray}
	Minimizing the total mean field energy leads to the self-consistent equations,
	\begin{eqnarray}
		(E_\Gamma+\lambda_{i})\phi_{i\Gamma}&=&\sum_{\alpha\sigma}\sum_{\Gamma'}
		\epsilon_{i\alpha\sigma}{\rm Re}[ n_{\alpha\sigma}^{\Gamma\Gamma'}]\phi_{i\Gamma'}-K_{i\Gamma}  \nonumber \\
		\sum_{\Gamma}\phi_{i\Gamma}^2&=&1 \nonumber \\
		\langle f_{i\alpha\sigma}^\dagger f_{i\alpha\sigma} \rangle&=&\sum_{\Gamma\Gamma'}{\rm Re} [n_{\alpha\sigma}^{\Gamma\Gamma'}]\phi_{i\Gamma}\phi_{i\Gamma'} \nonumber
	\end{eqnarray}
	where
	\begin{eqnarray}
		K_{i\Gamma}=\sum_{\alpha\sigma}\frac{\partial z_{i\alpha\sigma}}{\partial \phi_{i\Gamma}}\sum_{j\beta}z_{j\beta\sigma}t_{i\alpha,j\beta}{\rm Re}\langle f_{i\alpha\sigma}^\dagger f_{j\beta\sigma} \rangle. \nonumber
	\end{eqnarray}
	
	\subsection{Derivation of the Kugel-Khomskii superexchange model for the general two-orbital Hubbard model}
	
	For a single-orbital (or single-band) half-filled Hubbard model, the low energy physics in the strong coupling large-$U$ limit is described by the Heisenberg spin model with the superexchange interactions, since the charge excitations are gapped out. In the case of two degenerate orbitals, such as one electron in the degenerate $e_g$ orbitals, Kugel and Khomskii \cite{Skk1,Skk2} showed that the low energy physics in the strong-coupling limit corresponds to a superexchange model known as the Kugel-Khomskii model with both spin and orbital degrees of freedom. Here we first derive such a spin-orbital superexchange model for the general case with both orbital-dependent hopping and Hund's rule coupling, 
	which is necessary for studying the $e_g$ orbitals in the cuprates. We will then study the effects of the spin-orbital superexchange on the spin-singlet pairing.
	
	We start with the nearest neighbor tight-binding part
	\begin{eqnarray}
		H_{t}&=&\sum_{<ij>\alpha\beta\sigma}t_{\alpha\beta}
		d_{i\alpha\sigma}^\dagger d_{j\beta\sigma}+h.c.
	\end{eqnarray}
	where $\alpha,\beta=1,2$ denote the $d_{x^2-y^2}$ and $d_{z^2}$ orbitals of a general two-orbital system and $t_{\alpha\beta}$ describe the three independent parameters $t_{11}$, $t_{22}$, and $t_{12}$ for the intra and inter orbital hopping. The correlation part is given in the multi-orbital Hubbard model in Eq.~(\ref{hub}).
	To derive the effective low-energy theory in the strong coupling limit, it is convenient to consider the
	local Hilbert space at every site, which contains 16 states: an empty state, 4 singly-occupied states, and 11 excited states involving multiple occupations that are higher in energy by at least the order of $U$. The Hamiltonian matrix for a pair of two nearest neighbor singly-occupied sites $(i,j)$ has therefore the following form in the basis of the local Fock states
	\begin{eqnarray}
		\begin{bmatrix}
			H_{00} & H_{01} \\
			H_{10} & H_U
		\end{bmatrix}
	\end{eqnarray}
	where $H_{00}$ is a $16\times16$ matrix of the on-site energies spanned by the $16$ singly-occupied states at $i$ and $j$; $H_{01}$ is a $16\times12$ hopping matrix connecting the 16 ground states to the 12 excited states with one site empty and the other doubly occupied with $H_{10}=H_{01}^\dagger$; and $H_U$ is a $12\times12$ matrix of the excited states energies due to $\hH_I$. Projecting into the space of single occupation, we obtain the down-folded low-energy effective Hamiltonian
	\begin{eqnarray}
		\hH_{eff}\approx H_{00}-H_{01}H_{U}^{-1}H_{10}.
	\end{eqnarray}
	
	The remaining task is to write the second term in the above equation in terms of the spin and orbital operators. The matrix elements in $H_{01}$ and $H_{10}$ originate from the hopping between sites $i$ and $j$. Thus, this term has the most general form of $d_{i\alpha\sigma}^\dagger d_{i\beta'\sigma'}d_{j\alpha'\sigma'}^\dagger d_{j\beta\sigma}$, i.e. a product of bilinears at $i$ and $j$. In the spin-orbital basis $(d_{1\uparrow}, d_{1\downarrow},d_{2\uparrow}, d_{2\downarrow})^{T}$, the bilinear at each site can be represented by a tensor product
	$T_{i\mu}S_{i\nu}$ \cite{Skk1,Skk2}, where $S_{i\mu}$, $\mu=0,x,y,z$ are one half of the Pauli matrices acting in the spin space, while $T_{i\mu}$, $\mu=0,x,y,z$ are one half of the Pauli matrices acting in the orbital space.
	For example, it is easy to check that $d_{i1\uparrow}^\dagger d_{i1\uparrow}=(\frac{1}{2}+T_{iz})(\frac{1}{2}+S_{iz})$, $d_{i1\uparrow}^\dagger d_{i1\downarrow}=(\frac{1}{2}+T_{iz})S_{i}^{+}$ and $d_{i1\downarrow}^\dagger d_{i2\uparrow}=T_{i}^{+}S_{i}^{-}$, etc., where $T_{i}^{\pm}=T_{ix}\pm T_{iy}$ and $S_{i}^{\pm}=S_{ix}\pm S_{iy}$. 
	Following this procedure, we obtain the spin-orbital superexchange Hamiltonian in terms of the spin and orbital operators. Due to the spin rotation symmetry, $\hH_{eff}$ can be written as the sum of two contributions in the spin-singlet and spin triplet channels,
	\begin{eqnarray}
		\hH_{eff}&=&\sum_{\langle ij\rangle}-(\frac{1}{4}-{\bf S}_{i}\cdot {\bf S}_{j})\bigl[ \frac{t_{11}^2+2t_{12}^2+t_{22}^2}{2}(\frac{1}{U-J_H}+\frac{1}{U+J_H}
		+\frac{1}{U'+J_H}) \nonumber \\
		&& +t_{12}(t_{11}+t_{22})(\frac{1}{U+J_H}+\frac{1}{U'+J_H})
		(T_{i}^{+}+T_{i}^{-}+T_{j}^{+}+T_{j}^{-})+(t_{11}^2-t_{22}^2)
		(\frac{1}{U+J_H}+\frac{1}{U-J_H})(T_{iz}+T_{jz}) \nonumber \\ &&+(\frac{2t_{11}t_{22}}{U+J_H}-\frac{2t_{11}t_{22}}{U-J_H}+
		\frac{2t_{12}^2}{U'+J_H})(T_{i}^{+}T_{j}^{+}+T_{i}^{-}T_{j}^{-})+
		(\frac{2t_{12}^2}{U+J_H}-\frac{2t_{12}^2}{U-J_H}+\frac{2t_{11}t_{22}}
		{U'+J_H})(T_{i}^{+}T_{j}^{-}+T_{i}^{-}T_{j}^{+}) \nonumber \\
		&&+\frac{2t_{12}(t_{11}-t_{22})}{U+J_H}((T_{i}^{+}+T_{i}^{-})
		T_{jz}+T_{iz}(T_{j}^{+}+T_{j}^{-})) +2(t_{11}^2-2t_{12}^2+t_{22}^2)(\frac{1}{U-J_H}+\frac{1}
		{U+J_H}-\frac{1}{U'+J_H}) T_{iz}T_{jz}\nonumber \bigr]\\
		&&-(\frac{3}{4}+{\bf S}_{i}\cdot {\bf S}_{j})\bigl[  \frac{t_{11}^2+2t_{12}^2+t_{22}^2}{2(U'-J_H)}+2\frac{t_{11}^2-2t_{12}^2+t_{22}^2}{J_H-U'}T_{iz}T_{jz} \nonumber \\
		&& +\frac{2t_{12}^2}{J_H-U'}(T_{i}^{+}T_{j}^{+}+T_{i}^{-}T_{j}^{-})+\frac{2t_{11}t_{22}}{J_H-U'}(T_{i}^{+}T_{j}^{-}+T_{i}^{-}T_{j}^{+}) +\frac{2t_{12}(t_{11}-t_{22})}{J_H-U'}((T_{i}^{+}+T_{i}^{-})T_{jz}
		+T_{iz}(T_{j}^{+}+T_{j}^{-})) \bigr].  \nonumber
	\end{eqnarray}
	
	It is instructive to rewrite the above results in terms of exchange interactions in the spin, orbital, and spin-orbital sectors given in Eq.~(3) in the main text,
	\begin{eqnarray}
		\hH_{eff}=\hH_{\rm J-K}=\sum_{\langle ij\rangle}\biggl[ J{\bf S}_{i}\cdot{\bf S}_j
		&+&\sum_{\mu\nu} I_{\mu\nu} T_i^\mu T_j^\nu
		+\sum_{\mu\nu} K_{\mu\nu}({\bf S}_{i}\cdot{\bf S}_j)(T_i^\mu T_j^\nu)  \biggr],
		\label{hjk}
	\end{eqnarray}
	where the spin exchange coupling is given by
	\begin{equation}
		J=\frac{t_{11}^2+2t_{12}^2+t_{22}^2}{2}(\frac{1}{U-J_H}
		+\frac{1}{U+J_H}+\frac{1}{U'+J_H}-\frac{1}{U'-J_H}).
		\label{exch-j}
	\end{equation}
	The orbital exchange couplings are
	\begin{eqnarray}
		I_{00}&=&-\frac{t_{11}^2+2t_{12}^2+t_{22}^2}{2}(\frac{1}{4}(\frac{1}{U-J_H}+\frac{1}{U+J_H}+\frac{1}{U'+J_H})  +\frac{3}{4}\frac{1}{U'-J_H})   \nonumber \\
		I_{+0}&=&I_{-0}=I_{0+}=I_{0-}=-\frac{1}{4}t_{12}(t_{11}+t_{22})(\frac{1}{U+J_H}+\frac{1}{U'+J_H})  \nonumber \\
		I_{z0}&=&I_{0z}=-\frac{1}{4}(t_{11}^2-t_{22}^2)(\frac{1}{U+J_H}+\frac{1}{U-J_H})  \nonumber \\
		I_{zz}&=&-2(t_{11}^2-2t_{12}^2+t_{22}^2)(\frac{1}{4}(\frac{1}{U-J_H}+\frac{1}{U+J_H}-\frac{1}{U'+J_H})  -\frac{3}{4}\frac{1}{U'-J_H})   \nonumber \\
		I_{++}&=&I_{--}=-\frac{1}{4}(2t_{11}t_{22}(\frac{1}{U+J_H}-\frac{1}{U-J_H})+\frac{2t_{12}^2}{U'+J_H})  +\frac{3}{4}\frac{2t_{12}^2}{U'-J_H}   \nonumber \\
		I_{+-}&=&I_{-+}=-\frac{1}{4}(2t_{12}^2(\frac{1}{U+J_H}-\frac{1}{U-J_H})+\frac{2t_{11}t_{22}}{U'+J_H}) +\frac{3}{4}\frac{2t_{11}t_{22}}{U'-J_H}   \nonumber \\
		I_{+z}&=&I_{-z}=I_{z+}=-2t_{12}(t_{11}-t_{22})(\frac{1}{4}
		\frac{1}{U+J_H}-\frac{3}{4}\frac{1}{U'-J_H}).
		\label{exch-i}
	\end{eqnarray}
	The spin-orbital entangled coupling constants are given by
	\begin{eqnarray}
		K_{+0}&=&K_{-0}=K_{0+}=K_{0-}=t_{12}(t_{11}+t_{22})(\frac{1}{U+J_H}+\frac{1}{U'+J_H})  \nonumber \\
		K_{z0}&=&K_{0z}=(t_{11}^2-t_{22}^2)(\frac{1}{U+J_H}+\frac{1}{U-J_H})  \nonumber \\
		K_{zz}&=&2(t_{11}^2-2t_{12}^2+t_{22}^2)(\frac{1}{U-J_H}+\frac{1}{U+J_H}-\frac{1}{U'+J_H} +\frac{1}{U'-J_H})   \nonumber \\
		K_{++}&=&K_{--}=2t_{11}t_{22}(\frac{1}{U+J_H}-\frac{1}{U-J_H})+\frac{2t_{12}^2}{U'+J_H}  +\frac{2t_{12}^2}{U'-J_H}   \nonumber \\
		K_{+-}&=&K_{-+}=2t_{12}^2(\frac{1}{U+J_H}-\frac{1}{U-J_H})+\frac{2t_{11}t_{22}}{U'+J_H}  +\frac{2t_{11}t_{22}}{U'-J_H}   \nonumber \\
		K_{+z}&=&K_{-z}=K_{z+}=2t_{12}(t_{11}-t_{22})(\frac{1}{U+J_H}
		+\frac{1}{U'-J_H}).
		\label{exch-k}
	\end{eqnarray}
	
	Having derived the spin-orbital superexchange model given by Eqs.~(\ref{hjk}-\ref{exch-k}) for a general two-orbital system in the strong-coupling limit, we now turn to the case of the monolayer cuprates in the hole-rich regime. Since the $d_{x^2-y^2}$ and $d_{z^2}$ orbitals are still significantly split by the Jahn-Teller distortion, orbital order takes place and the operator $T_{iz}$ can be replaced by its expectation value $\frac{1}{2}$ and arrive at,
	\begin{eqnarray}
		\hH_{\rm J-K}&=&\sum_{\langle ij\rangle}\biggl[J_s{\bf S}_{i}\cdot{\bf S}_j+I_0+\sum_{\mu\nu} (I_s+K_s({\bf S}_{i}\cdot{\bf S}_j))(T_i^+ +T_i^- +T_j^+ +T_j^-)
		\nonumber \\
		&+&(I+K({\bf S}_{i}\cdot{\bf S}_j))(T_i^+ T_j^++T_i^+ T_j^-+T_i^- T_j^++T_i^- T_j^-)
		\biggr],
		\label{hjk-quench}
	\end{eqnarray}
	where $J_s=J+K_{0z}+\frac{K_{zz}}{4}$, $I_0=I_{00}+I_{0z}+\frac{I_{zz}}{4}$, $I=I_{++}$, $K=K_{++}$, $K_s=K_{+0}+\frac{K_{+z}}{2}$, and $I_s=I_{+0}+\frac{I_{+z}}{2}$ are combinations of the exchange couplings given in Eqs.(\ref{exch-i}) and (\ref{exch-k}). It is instructive to verify that in the limit $J_H=0$, $U=U^\prime$ and the first term in Eq.~(\ref{hjk-quench}) reduces to the Heisenberg interaction in the single-band $t$-$J$ model with $J_s=4t_{11}^2/U$.
	
	However, it is important to realize that $T_{iz}$ order does not quench entirely the orbital degrees of freedom. The transverse orbital fluctuations represented by $T_i^{\pm}$ in Eq.~(\ref{hjk-quench}) make important contributions to pairing. Indeed, the exchange interactions in Eq.~(\ref{hjk-quench}) can be written in terms spin-singlet pairing,
	\begin{eqnarray}
		\hH_{\rm J-K}^\Delta&=&-\frac{1}{2}\sum_{\langle ij \rangle}\biggl[ J_s(\Delta_{ij}^{11\dagger}\Delta_{ij}^{11}
		+\Delta_{ij}^{22\dagger}\Delta_{ij}^{22}+\Delta_{ij}^{12\dagger}\Delta_{ij}^{12}+\Delta_{ij}^{21\dagger}\Delta_{ij}^{21})  \nonumber \\
		&& +K(\Delta_{ij}^{11\dagger}\Delta_{ij}^{22}+\Delta_{ij}^{22\dagger}\Delta_{ij}^{11}+\Delta_{ij}^{12\dagger}\Delta_{ij}^{21}+\Delta_{ij}^{21\dagger}\Delta_{ij}^{12})  \nonumber \\
		&& +K_s(\Delta_{ij}^{11\dagger}\Delta_{ij}^{12}+\Delta_{ij}^{12\dagger}\Delta_{ij}^{21}+\Delta_{ij}^{22\dagger}\Delta_{ij}^{12}+\Delta_{ij}^{22\dagger}\Delta_{ij}^{21}+h.c.) \biggr],  \label{jkd}
	\end{eqnarray}
	where we have introduced the pairing order parameters,
	\begin{eqnarray}
		\Delta_{ij}^{\alpha\beta\dagger}&=&d_{i\alpha\uparrow}^\dagger d_{j\beta\downarrow}^\dagger-d_{i\alpha\downarrow}^\dagger  d_{j\beta\uparrow}^\dagger\nonumber \\ \Delta_{ij}^{\alpha\beta}&=&d_{j\beta\downarrow}d_{i\alpha\uparrow}
		-d_{j\beta\uparrow}d_{i\alpha\downarrow}
	\end{eqnarray}
	and ignored constants. It is now manifest that the spin-orbital exchange interaction $K$ gives rise to the important interorbital scattering of pairs formed in the individual orbitals, i.e. terms like $-K\Delta_{ij}^{11\dagger}\Delta_{ij}^{22}$ etc. These terms are the strong coupling counterparts of the inter FS pocket pair scattering in the weak-coupling approaches for multiband superconductivity. In the strong-coupling Gutzwiller calculation, we insert Eq.~(\ref{jkd}) into the Eq.~(\ref{h-gutzwiller}), decouple the pairing interactions at the meanfield level by the average pairing fields
	\begin{eqnarray}
		\Delta_{ij}^{\alpha\beta\dagger}\Delta_{ij}^{\alpha'\beta'}
		\approx\la\Delta_{ij}^{\alpha\beta\dagger}
		\ra\Delta_{ij}^{\alpha'\beta'}+\Delta_{ij}^{\alpha\beta\dagger}
		\la\Delta_{ij}^{\alpha'\beta'}\ra-\la\Delta_{ij}^{\alpha\beta\dagger}
		\ra\la\Delta_{ij}^{\alpha'\beta'}\ra,
	\end{eqnarray}
	and determine their expectation values self-consistently using Eq.~(\ref{pp}). In general, the structure of the pairing fields should obey the crystal symmetry and thus has the form
	\begin{eqnarray}
		\langle \Delta_{ij}^{\alpha\beta}\rangle&=&\frac{1}{N_s}
		\sum_{\mathbf{k},\alpha\beta}\Delta_{\alpha\beta}
		b_{\alpha\beta}(\mathbf{k})e^{i\mathbf{k}(r_i-r_j)},
		\label{pairing_s}
	\end{eqnarray}
	where $N_s$ is the number of lattice sites and $b_{\alpha\beta}(k)$ the form factors of different symmetries in the $C_{4v}$ point group including the $A_1$ and $B_1$ channels. For the $A_1$ channel,
	\begin{eqnarray}
		b_{11/22}&=&\cos k_x+\cos k_y \nonumber \\
		b_{12/21}&=&\cos k_x-\cos k_y.
		\label{A1}
	\end{eqnarray}
	For the $B_1$ channel,
	\begin{eqnarray}
		b_{11/22}&=&\cos k_x-\cos k_y  \nonumber \\
		b_{12/21}&=&\cos k_x+\cos k_y.
		\label{B1}
	\end{eqnarray}
	We find that in the hole-rich regime with $0.7\le x_h\le 0.9$, the pairing is dominated by intra-orbital pairing in the $A_1$ channel, i.e. by extended $s$-wave pairing. Moreover, it can be seen from Eq.~(\ref{jkd}) that the pairing interaction induced by $K_s$ mixes the $A_1$ and $B_1$ channels, corresponding to $s$-$d$ mixed pairing that is negligibly small in the hole-rich regime. Thus, in the regime of our interests, $K_s$ can be ignored for the analysis in self-consistent solutions of purely $A_1$ and $B_1$ symmetry. At this stage, the exchange couplings are treated as phenomenological parameters. We set $J_s=120$meV, a commonly accepted value for the bulk cuprates, $K_s=0$, and study two cases with $K=60$meV and $80$meV respectively. The corresponding tunneling density of states in the resulting nodeless SC state at $x_h=0.9$ and $n_e=2.1$ is shown in Figs 3(c) and 3(d) in the main text.
	
\end{widetext}

\end{document}